\documentclass[12pt]{iopart}

\usepackage{iopams}  
\usepackage{graphicx}

\begin{document}

\title[Reply to Comment on "Quantum inversion of cold atoms in a
 ..." ]
{\bf Reply to Comment on "Quantum inversion of cold atoms in a
microcavity: spatial dependence"
}

\author[Abdel-Aty and Obada]{Mahmoud Abdel-Aty$^{1,}$\footnote[3]{E-mail: abdelatyquant@yahoo.co.uk}, A.-S. F. Obada$^2$ 
}
\address{$^1$Mathematics Department, Faculty of Science, South Valley University, 82524 Sohag, Egypt.
\\
$^2$Mathematics Department, Faculty of Science, Al-Azhar University, Nasr City, Cairo, Egypt.
}

\begin{abstract}

The question raised by [Bastin and Martin 2003 J. Phys. B: At. Mol. Opt. Phys. {\bf 36}, 4201] is examined and used to
explain in more detail a key point of our calculations. They have sought to
rebut criticisms raised by us of certain techniques used in the calculation
of the off-resonance case. It is also explained why this result is not a
problem for the off-resonance case, but, on the contrary, opens the door to
a general situation. Their comment is based on a blatant misunderstanding of
our proposal an d as such is simply wrong.
\end{abstract}

{\bf Submitted to:} {\JPB}

\maketitle

In our paper [2] we have used the Hamiltonian (in the mesa mode case)

\begin{equation}
\hat{H}=\frac{p_{z}^{2}}{2M}+\frac{\Delta }{2}\sigma_{z}+\omega (a^{\dagger
}a+\frac{1}{2}\sigma_{z})+\lambda f(z)\{\sigma ^{-}a^{\dagger }+a\sigma
^{+}\}.  \label{1}
\end{equation}
Let us write equation (1) in the following form (in the mesa mode case $%
f(z)=1)$
\begin{eqnarray}
\hat{H} &=&\frac{p_{z}^{2}}{2M}+\hat{V}  
\nonumber 
\\
\hat{V} &=&\frac{\Delta }{2}\sigma_{z}+\omega (a^{\dagger }a+\frac{1}{2}%
\sigma_{z})+\lambda \{\sigma ^{-}a^{\dagger }+a\sigma ^{+}\}.
\end{eqnarray}
It is easy to show that in the $2\times 2$ atomic-photon space the
eigenvalues and eigenfunction of the interaction Hamiltonian $\hat{V}$ ($%
\hat{V}|\Phi_{n}^{\pm }\rangle =E_{n}^{\pm }|\Phi_{n}^{\pm }\rangle )$ are
given by [3],\smallskip 
\begin{equation}
E_{n}^{\pm }=\omega (n+\frac{1}{2})\pm \sqrt{\frac{\Delta ^{2}}{4}+\lambda
^{2}(n+1)},
\end{equation}
\begin{eqnarray}
|\Phi_{n}^{+}\rangle &=&\cos \theta_{n}|n+1,g\rangle +\sin \theta
_{n}|n,e\rangle,  
\nonumber 
\\
|\Phi_{n}^{-}\rangle &=&-\sin \theta_{n}|n+1,g\rangle +\cos \theta
_{n}|n,e\rangle,
\end{eqnarray}
where 
\begin{equation}
\theta_{n}=\tan ^{-1}\left( \frac{\lambda \sqrt{n+1}}{\sqrt{\frac{\Delta
^{2}}{4}+\lambda ^{2}(n+1)}-\frac{\Delta }{2}}\right),\qquad 
\end{equation}
We write the wave function $|\Psi (z,t)\rangle =\sum\limits_{n}C_{n}^{\pm
}(z,t)|\Phi_{n}^{\pm }\rangle .$ Then using the total Hamiltonian (1) we
have 
\begin{eqnarray}
\hat{H}|\Psi (z,t)\rangle  &=&\sum\limits_{n}\left( \frac{p^{2}}{2M}%
+V\right) C_{n}^{\pm }(z,t)|\Phi_{n}^{\pm }\rangle   
\nonumber 
\\
&=&\sum\limits_{n}\frac{p^{2}}{2M}C_{n}^{\pm }(z,t)|\Phi_{n}^{\pm }\rangle
+\sum\limits_{n}VC_{n}^{\pm }(z,t)|\Phi_{n}^{\pm }\rangle   
\nonumber 
\\
&=&\sum\limits_{n}\frac{p^{2}}{2M}C_{n}^{\pm }(z,t)|\Phi_{n}^{\pm }\rangle
+\sum\limits_{n}C_{n}^{\pm }(z,t)E_{n}^{\pm }|\Phi_{n}^{\pm }\rangle  
\nonumber 
\\
&=&\sum\limits_{n}\left( \frac{p^{2}}{2M}C_{n}^{\pm }(z,t)+E_{n}^{\pm
}C_{n}^{\pm }(z,t)\right) |\Phi_{n}^{\pm }\rangle ,
\end{eqnarray}
because of the orthonormality of the wavefunctions $|\Phi_{n}^{\pm }\rangle $
then
\begin{equation}
i\frac{\partial }{\partial t}C_{n}^{\pm }=\left( -\frac{1}{2M}\frac{\partial
^{2}}{\partial z^{2}}+E_{n}^{\pm }\right) C_{n}^{\pm },
\end{equation}
with no coupling even in the presence of the detuning (equation (13) in BM
comment [1]).

It may be worthwhile for the authors to consult some papers that have been
published previously (see for example Refs. [4-6]) where the detuning has
been considered and similar results have been obtained. To be more precisely:

\begin{itemize}
\item[1-]  We have used the interaction picture, so that the term $(n+\frac{1%
}{2})\omega $ does not appear, it can be used as a phase only. Bearing in
mind the case of mesa mode is being treated in our paper i.e. $f(z)=1$.

\item[2-]  The most serious point is that Bastin and Martin have overlooked
the formulae for 
\begin{equation}
\cos 2\theta_{n}\qquad and\qquad \sin 2\theta_{n}.
\end{equation}
From equation (10) raised in Bastin and Martin comment, it is easy to write 
\begin{equation}
\tan \theta_{n}=\frac{\lambda \sqrt{n+1}}{\sqrt{\frac{\Delta ^{2}}{4}%
+\lambda ^{2}(n+1)}-\frac{\Delta }{2}}=\frac{\sqrt{\frac{\Delta ^{2}}{4}%
+\lambda ^{2}(n+1)}+\frac{\Delta }{2}}{\lambda \sqrt{n+1}}
\end{equation}
then 
\begin{equation}
\tan 2\theta_{n}=\frac{\lambda \sqrt{n+1}}{-\frac{\Delta }{2}}
\end{equation}
Also, it is easy to prove that 
\begin{equation}
\cos 2\theta_{n}=\frac{-\Delta /2}{\sqrt{\frac{\Delta ^{2}}{4}+\lambda
^{2}(n+1)}}\qquad and\qquad \sin 2\theta_{n}=\frac{\lambda \sqrt{n+1}}{%
\sqrt{\frac{\Delta ^{2}}{4}+\lambda ^{2}(n+1)}}.
\end{equation}
Once these formulae inserted in equations (18) and (19) of Bastin and Martin
comment, we find that, the second terms vanish identically.
\end{itemize}

Now let us look more carefully at the general case when we take f(z) no
longer a constant, i.e. we go beyond the mesa mode case. In this case the
orthonormal functions $|\Phi_{n}^{\pm }\rangle $ in the $2\times 2$ system
diagonalize the Hamiltonian $V$ and its elements are diagonal in this set
of functions with 
\begin{eqnarray}
V_{n}^{\pm } &=&(n+\frac{1}{2})\omega \pm \sqrt{\frac{\Delta ^{2}}{4}%
+\lambda ^{2}f^{2}(z)(n+1)}  \nonumber \\
\tan 2\theta_{n} &=&\frac{\lambda f(z)\sqrt{n+1}}{-\frac{\Delta }{2}}.
\end{eqnarray}
The states $|\Phi_{n}^{\pm }\rangle $ are z-dependent through the
trigonometric functions, they satisfy
\begin{eqnarray}
\frac{\partial }{\partial z}|\Phi_{n}^{\pm }\rangle &=&\pm |\Phi_{n}^{\pm
}\rangle \frac{d\theta }{dz},  \nonumber \\
\frac{\partial ^{2}}{\partial z^{2}}|\Phi_{n}^{\pm }\rangle &=&\pm |\Phi
_{n}^{\pm }\rangle \frac{d^{2}\theta }{dz^{2}}-|\Phi_{n}^{\pm }\rangle
\left( \frac{d\theta }{dz}\right) ^{2}.
\end{eqnarray}
Then $|\Psi (z,t)\rangle $ can be expanded in the form $|\Psi (z,t)\rangle
=\sum\limits_{n}C_{n}^{\pm }(z,t)|\Phi_{n}^{\pm }\rangle $ and it satisfies
the Schrodinger equation 
\begin{equation}
i\frac{\partial }{\partial z}|\Psi (z,t)\rangle =H|\Psi (z,t)\rangle .
\end{equation}

Hence the coefficients $C_{n}^{\pm }(z,t)$ satisfy the coupled equations 
\begin{eqnarray}
i\frac{\partial C_{n}^{+}}{\partial z} &=&\left( -\frac{1}{2M}\frac{\partial
^{2}}{\partial z^{2}}+V_{n}^{+}-\left( \frac{d\theta }{dz}\right)
^{2}\right) C_{n}^{+}-\left( 2\frac{\partial C_{n}^{-}}{\partial z}\left( 
\frac{d\theta }{dz}\right) +C_{n}^{-}\left( \frac{d\theta }{dz}\right)
^{2}\right) ,  
\nonumber 
\\
i\frac{\partial C_{n}^{-}}{\partial z} &=&-\left( -\frac{1}{2M}\frac{%
\partial ^{2}}{\partial z^{2}}+V_{n}^{-}-\left( \frac{d\theta }{dz}\right)
^{2}\right) C_{n}^{-}+\left( 2\frac{\partial C_{n}^{+}}{\partial z}\left( 
\frac{d\theta }{dz}\right) +C_{n}^{+}\left( \frac{d\theta }{dz}\right)
^{2}\right) ,
\end{eqnarray}
These equations should replace equations (18) and (19) of the comment of
[1]. But once $f(z)$ is taken to be constant, then $\frac{d\theta }{dz}$
will vanish and we get equations (7) and the results of [2,3].

\end{document}